\documentclass{aastex61}

\accepted{March 23, 2018}
\submitjournal{ApJ}

\shorttitle{Orbital Characteristics of the EC20117-4014 binary system}
\shortauthors{Otani et al.}

\begin{document}

\title{Orbital Characteristics of the Subdwarf-B and F V Star Binary EC~20117-4014(=V4640 Sgr)}

\correspondingauthor{T. Otani}
\email{otanit@erau.edu}

\author{T. Otani}
\affil{Department of Physical Sciences and SARA, Embry-Riddle Aeronautical University, 600 S. Clyde Morris Blvd, Daytona Beach, \\FL 32114, United States}
\affil{Department of Physics \& Space Sciences, Florida Institute of Technology, 150 W University Blvd, Melbourne, FL 32901, United States}

\author{T. D. Oswalt}
\affiliation{Department of Physical Sciences and SARA, Embry-Riddle Aeronautical University, 600 S. Clyde Morris Blvd, Daytona Beach, \\FL 32114, United States}

\author{A. E. Lynas-Gray}
\affiliation{Department of Physics, University of Oxford, Oxford, OX1 3RH, United Kingdom}

\author{D. Kilkenny}
\affiliation{Department of Physics and Astronomy, University of the Western
Cape, Bellville 7535, South Africa}

\author{C. Koen}
\affiliation{Department of Statistics, University of the Western
Cape, Bellville 7535, South Africa}

\author{M. Amaral}
\altaffiliation{Southeastern Association for Research in Astronomy (SARA) NSF-REU Summer Intern (2015)}
\affiliation{Department of Science, Cape Cod Community College, 2240 Iyannough Rd, West Barnstable, MA 02668, United States}

\author{R. Jordan}
\altaffiliation{Southeastern Association for Research in Astronomy (SARA) NSF-REU Summer Intern (2015)}
\affiliation{Astronomy Department, Whitman College, 345 Boyer Ave, Walla Walla, WA 99362, United States}



\begin{abstract}

Among the competing evolution theories for subdwarf-B (sdB) stars is the binary evolution scenario. EC~20117-4014 (=V4640~Sgr) is a spectroscopic binary system consisting of a pulsating sdB star and a late F main-sequence companion \citep{O'Donoghue et al.1997}, however the period and the orbit semi-major axes have not been precisely determined. This paper presents orbital characteristics of the EC 20117-4014
binary system using 20 years of photometric data. Periodic Observed minus Calculated (O-C) variations were detected in the two highest amplitude pulsations identified in the EC 20117-4014 power spectrum, indicating the binary system's precise orbital period (P = 792.3 days) and the light-travel time amplitude (A = 468.9 s).  This binary shows no significant orbital eccentricity and the upper limit of the eccentricity is 0.025 (using 3$\sigma$ as an upper limit).  This upper limit of the eccentricity is the lowest among all wide sdB binaries with known orbital parameters.  This analysis indicated that the sdB is likely to have lost its hydrogen envelope through stable Roche lobe overflow, thus supporting hypotheses for the origin of sdB stars.  In addition to those results, the underlying pulsation period change obtained from the photometric data was $\dot{P}$ = 5.4 ($\pm$0.7) $\times$ $10^{-14}$ d d$^{-1}$, which shows that the sdB is just before the end of the core helium-burning phase. 

\end{abstract}




\section{Introduction} \label{sec:intro}

      Subdwarf B (sdB) stars are core helium burning objects, found in both the disk and halo of our Galaxy \citep{Saffer et al.1994}. V361 Hya was the first pulsating sdB star to be discovered \citep{Kilkenny et al.1997} and \citet{Ostensen et al.2010} subsequently 
discovered only twenty new pulsators among the more than 300 sdBs predicted to lie in the V361 Hya instability strip and
monitored, suggesting only about 10$\%$ of these sdB stars have pulsations detectable from the ground.
The sdB observed properties place them in the extreme horizontal branch (EHB) part of the H-R diagram. Their effective temperatures range from 22,000 to 40,000 $K$ and surface gravities range from 5.0 $\le$ log $g$ $\le$6.2 (in cgs units).  Their masses are narrowly confined to about 0.5$M_{\sun}$ \citep{Heber2009}.  Subdwarf B stars have experienced mass-loss at the end of the red giant branch phase \citep{Bonanno et al.2003}, in which the hydrogen envelope is lost, leaving a helium core with a very thin inert hydrogen-rich envelope. The loss of the hydrogen envelope prevents the star from ascending the asymptotic giant branch and the star settles on the EHB, spending about 10$^{8}$ years as an sdB star.  Upon helium depletion in the core they become subdwarf O (sdO) stars burning helium in a shell surrounding a C/O core and, eventually, DAO white dwarfs \citep{Dorman et al.1993, Bergeron et al.1994}. 
      
      The formation mechanism of sdB stars, i.e. why they lose their hydrogen envelopes, is a matter of current debate.   Plausible sdB formation models via binary evolution were constructed by \citet{Han et al.2002, Han et al.2003}.  According to \citet{Silvotti et al.2011}, companion stars have been detected in at least 50$\%$ of sdB stars, strongly supporting a binary origin. However, some fraction of sdB stars may not be in binaries \citep{Heber2009, Fontaine et al.2012}.  If true, this would require another formation channel, perhaps the single star evolution scenario proposed by \citet{Dorman et al.1993}.  Their study of 105 single or wide-binary sdB stars showed that the binary evolution model of \citet{Han et al.2002, Han et al.2003} overestimates the number of sdB stars formed through the white dwarf merger channel. In another scenario, the merger of a helium white dwarf with a low-mass hydrogen burning star was proposed as a way of forming single sdB stars \citep{Clausen et al.2011}.  To distinguish between these evolutionary scenarios, orbital information on sdB star binaries is essential. 
      
      The existence of sdB pulsators (sdBV) was predicted by \citet{Charpinet et al.1996}.  Independently, \citet{Kilkenny et al.1997} discovered the first short period sdBV star, \objectname{EC 14026-2647}.  These stars are p-mode pulsators, where pulsations are driven by internal pressure fluctuations \citep{Charpinet et al.2000}.  The first long period sdBV star, \objectname{PG 1716+426}, is a g-mode pulsator \citep{Green et al.2003}, in which gravity provides the restoring force.  Some sdB stars have been discovered to exhibit both p-mode pulsations and g-mode pulsations.  These objects are called hybrid pulsators \citep{Schuh et al.2006, Oreiro et al.2004}. 
      
      The pulsation periods of sdBV stars are usually stable \citep{Ostensen et al.2001}, and therefore they are good chronometers.  A star's position in space may wobble due to the gravitational perturbations of a companion.  From an observer's point of view the light from the pulsating star is periodically delayed when it is on the far side of its orbit and advanced on the near side. Changes in the pulse arrival times are detected using the observed-minus-calculated (O-C) diagram. The O-C diagram is a technique that has long been used in the binary star community to search for additional components, orbital period changes, mass loss, etc.
      
      Several planets and substellar companions to sdB host stars have been detected by this method. \citet{Silvotti et al.2007} were the first to detect a planet around the sdB star \objectname{V391 Peg} in this way.  \citet{Lutz2011} detected companions to the sdB stars \objectname{HS 0444+0458} and \objectname{HS 0702+6043} which appear to be a brown dwarf and an exoplanet, respectively.  \citet{Mullally et al.2008} used the O-C method to search for possible planets around DAV white dwarfs.  Among the 15 white dwarf stars they surveyed, GD 66 exhibited O-C variations consistent with a 2 $M_{J}$ planet in a 4.5 year orbit.  Also, several planets and companion stars to sdB host stars have been detected by the O-C method using eclipse timings.  A companion to the sdB star HS 0705+6700 was similarly detected by \citet{Qian et al.2009}.        
      While several authors have reported the existence of planets orbiting
post-common envelope binaries, these conclusions must be regarded
as tentative because they are based 
on data obtained over a time interval which is the
same order as the orbital period proposed for an orbiting planet;
further observations over several of the proposed orbital periods
would be needed for confirmation. 
\citet{1992ApJ...385..621A} proposed an alternative explanation for
eclipse timing variations, often referred to as the 
``Applegate mechanism'', as a gravitational coupling of the orbit to
changes in the shape of a magnetically active star in the system.
\citet{2013A&A...549A..95Z} argued for additional observations to
distinguish between the Applegate mechanism and planet formation
following a common envelope phase. Planets may also survive 
a common envelope phase of their binary star host, as 
\citet{2016ApJ...832..183K} and \citet{2017MNRAS.465.2053V} discuss.
\citet{2016A&A...587A..34V} concluded that an improved version of
the Applegate mechanism, which includes angular momentum exchange
between a finite shell and the stellar core, cannot uniquely explain
orbital period variations in the sixteen systems they consider.
A further possibility comes from calculations by 
\citet{2017ApJ...837L..19C} which indicate that observed orbital period
derivatives in two post-common envelope binaries involving a hot
subdwarf (HW~Vir and NY~Vir) could
be produced by a resonant interaction between the binary and a 
circumbinary disk having a mass in the range $10^{-4}{\,}{\rm M}_{\odot}
- 10^{-2}{\,}{\rm M}_{\odot}$.
      
      \objectname[V4640 Sgr]{EC 20117-4014} (=V4640 Sgr) is one of the first p-mode sdBV stars to be discovered, and it is known to be a spectroscopic binary system \citep{O'Donoghue et al.1997}.  This star was also selected by \citet{Otani2015} as part of a three-year observational search for substellar companions among known sdB pulsators using the O-C method.  That project had two goals: (1) determine whether the frequency of detectable companions supports the hypothesis that all, or nearly all, were formed via binary interactions and (2) to examine the frequency with which post main sequence stars in general might harbor ``planetary survivors". So far, we have a total of about 20 years of EC 20117-4014 photometric data. In this paper, we present the results of our O-C analysis, and orbital information on the sdB and late F main-sequence companion binary system.
      
      Section~\ref{sec:obser} provides a summary of the physical properties of \objectname[V4640 Sgr]{EC 20117-4014} and a discussion of the facilities and instrumentation used to obtain the data needed for the O-C analysis.  Section~\ref{sec:OCmethod} outlines our reduction and analysis procedures.  Section~\ref{sec:results} presents our results derived from the observed pulsation peaks in the frequency spectrum of \objectname[V4640 Sgr]{EC 20117-4014}, how they have been used to derive useful constraints on a previously-reported F5 Main Sequence companion.  Also, the star's evolutionary phase is discussed.  Our conclusions and suggested additional work are summarised in Section~\ref{sec:conc}.

\section{Target and Observations} \label{sec:obser}

\objectname[V4640 Sgr]{EC 20117-4014} (=V4640 Sgr) is an sdB star originally found in the Edinburgh-Cape (EC) Blue Object Survey \citep{Stobie et al.1997, Kilkenny et al.2016}. The composite spectra obtained by \citet{O'Donoghue et al.1997} suggested that the sdB component of the \objectname[V4640 Sgr]{EC 20117-4014}  binary has a late F main-sequence companion.  This companion contributes more than half of the observed flux in the visual band. The apparent magnitude of the system is V = 12.47 $\pm$ 0.01, and the magnitude of the sdB star itself is V=13.55 $\pm$0.05 \citep{O'Donoghue et al.1997}. Three pulsation frequencies (7.29 mHz (137.3 s), 7.04 mHz (142.0 s) and 6.35 mHz (157.5 s)) were detected by \citet{O'Donoghue et al.1997}, making \objectname[V4640 Sgr]{EC 20117-4014} the fourth member of the class of short period sdBV stars.  Further high speed photometry of \objectname[V4640 Sgr]{EC 20117-4014} was obtained and an asteroseismology analysis was performed by \citet{Randall et al.2006}, who estimated the effective temperature (32,800 $\le$ $T_{\textrm{eff}}$ $\le$  36,800 K), the gravity (5.848 $\le$ $\log{g}$ $\le$ 5.864), and the mass (0.50 $\le$ $M_{\sun}$ $\le$ 0.59).   According to \citet{Randall et al.2006}, the companion is confirmed to be a late F main-sequence star, but they were unable to set strong constraints on its orbital period.  \citet{Lynas-Gray2013} found that the largest pulsation frequency (7.29 mHz) exhibits day-to-day amplitude changes.  Other sdBV stars, V541 Hya and KIC 010139564, show similar pulsation amplitude changes that are explained by rotational splitting \citep{Baran et al.2012, Randall et al.2009}.  This suggests the pulsation amplitude changes of \objectname[V4640 Sgr]{EC 20117-4014} may be a consequence of unresolved rotational splitting. The largest pulsation frequency amplitudes in the 2010-2011 data obtained by \citet{Otani2015} showed a linear decrease with time. 
      
\begin{figure}
\plotone{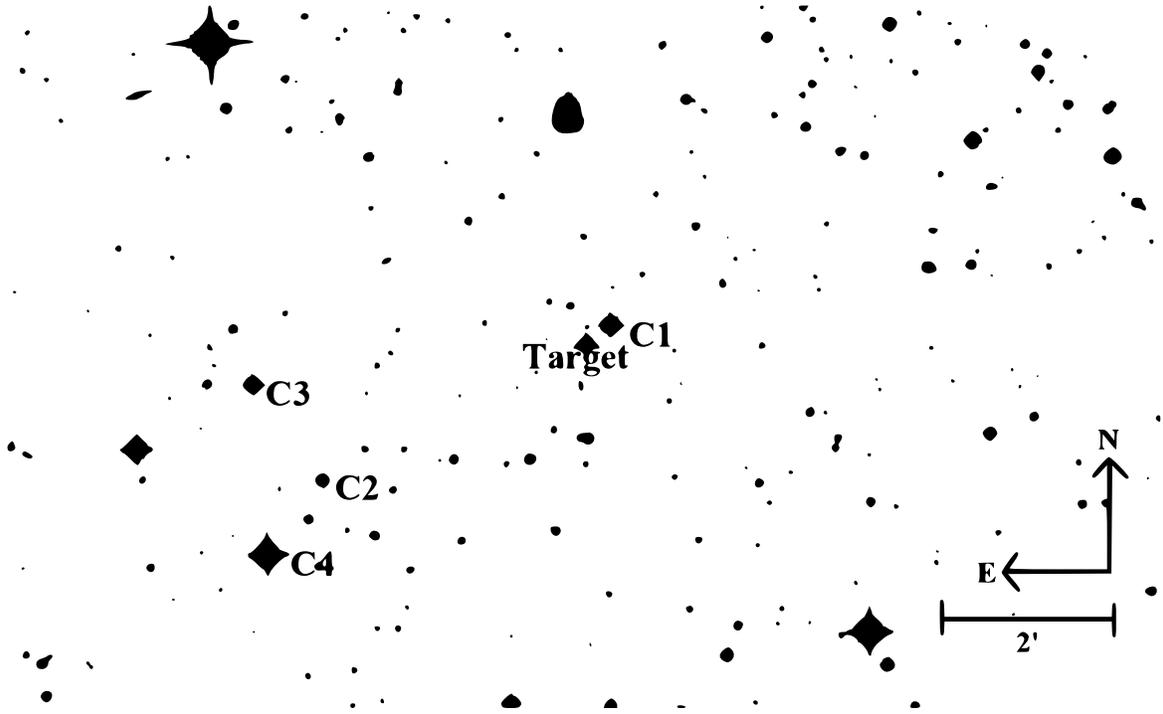}
\caption{Finder Chart of EC20117-4014 (http://simbad.u-strasbg.fr). The target and the comparison stars (C1-C4) are shown.\label{fig:figure1}}
\end{figure}

      The discovery observations of \objectname[V4640 Sgr]{EC 20117-4014} were made with the 0.75-m
and 1.0-m telescopes at the Sutherland site of the South African
Astronomical Observatory (SAAO) in 1995 (see Table 1 of \citet{O'Donoghue et al.1997}). At that time, both telescopes had (S-11) photomultiplier-based
photometers (the response of the system has a similar effective wavelength to the Johnson B, but with a much broader bandpass), though these were replaced by the UCTCCD photometer
(e.g. \citet{O'Donoghue et al.1996}) in the late 1990s. The UCTCCD
could be used on either telescope and our 0.75-m and 1.0-m data post-1996
were obtained with the UCTCCD. Because \objectname[V4640 Sgr]{EC 20117-4014} is relatively bright
it was mainly monitored with the SAAO 0.5m telescope between the discovery
in 1995 and 2000. The 0.5-m was permanently equipped with a (GaAs) photomultiplier-based photometer until recently decomissioned.  For the 0.75-m SAAO telescope observations in 2001 and 2011, UCTCCD was used with a Johnson-B filter.  Since
2010, EC~20117-4014 has also been monitored using the SARA-CT 0.6m telescope
at the Cerro Tololo InterAmerican Observatory (CTIO) in Chile\footnote{The SARA Observatory 0.6-m telescope at Cerro Tololo, Chile, is owned and operated by the Southeastern Association for Research in Astronomy ({\url{saraobservatory.org}}).}. Figure~\ref{fig:figure1}
provides a finder chart for the target and comparison stars.  For the SAAO CCD data, the only comparison star on the CCD chip was GSC07952-01358 (C1 in Table~\ref{tab:table1}).  Therefore, this star was used as a comparison star for the SAAO CCD observations. For the SARA-CT observations, we used four comparison stars (including GSC07952-01358). The coordinates of these target and comparison stars are given in Table~\ref{tab:table1}.  The observation log is presented in Table~\ref{tab:table2}.  

\begin{deluxetable}{ccc}
\tablecaption{Coordinates of the Target and Comparison Stars (epoch 2000) (http://simbad.u-strasbg.fr)\label{tab:table1}}
\tablenum{1}
\tablehead{\colhead{Identification Name} & \colhead{RA} & \colhead{DEC} \\ 
\colhead{} & \colhead{} & \colhead{} } 
\startdata
Target & 20$^{h}$15$^{m}$04.$^{s}$79 & -40\arcdeg 05'44.2" \\
C1 & 20$^{h}$15$^{m}$03.$^{s}$23 & -40\arcdeg 05'29.1" \\
C2 & 20$^{h}$15$^{m}$22.$^{s}$20 & -40\arcdeg 07'25.2" \\
C3 & 20$^{h}$15$^{m}$26.$^{s}$68 & -40\arcdeg 06'12.6" \\
C4 & 20$^{h}$15$^{m}$25.$^{s}$91 & -40\arcdeg 08'22.4" \\
\enddata
\tablecomments{For SAAO observations, only C1 (GSC 07952-01358) was used since this is the only comparison star in the field.}
\end{deluxetable}

\startlongtable
\begin{deluxetable}{lcccc}
\tablecaption{Observation Log for EC20117-4014\label{tab:table2}}
\tablenum{2}
\tablehead{\colhead{Date} & \colhead{Start of Run} & \colhead{Length} & \colhead{Observatory} & \colhead{Publications using the data} \\ 
\colhead{} & \colhead{(BJD-2445000)} & \colhead{(hr)} & \colhead{} & \colhead{} } 
\startdata
1995 May 31 & 4869.49293 & 4.4 & SAAO & 1 \\
1995 Aug 17 & 4947.23604 & 8.0 & SAAO & 1 \\
1995 Aug 18 & 4948.23415 & 2.7 & SAAO & 1 \\
1995 Aug 25 & 4955.32857 & 4.1 & SAAO & 1 \\
1995 Aug 27 & 4957.25274 & 5.8 & SAAO & 1 \\
1995 Sep 19 & 4980.26625 & 4.3 & SAAO & 1 \\
1996 Mar 26 & 5168.58856 & 1.4 & SAAO &  \\
1996 Apr 21 & 5194.61163 & 1.0 & SAAO &  \\
1996 Apr 22 & 5195.60918 & 1.3 & SAAO &  \\
1996 May 26 & 5229.55177 & 1.5 & SAAO &  \\
1996 May 27 & 5230.55181 & 1.5 & SAAO &  \\
1996 Jun 28 & 5262.53746 & 1.0 & SAAO & \\
1996 Jul 7 & 5272.48776 & 1.1 & SAAO & \\
1996 Jul 11 & 5276.46623 & 1.1 & SAAO & \\
1996 Jul 16 & 5281.47772 & 1.0 & SAAO & \\
1996 Jul 20 & 5285.45608 & 1.0 & SAAO & \\
1996 Aug 20 & 5316.44773 & 1.1 & SAAO & \\
1996 Aug 21 & 5317.47060 & 1.1 & SAAO & \\
1996 Sep 4 & 5331.41769 & 1.6 & SAAO & \\
1996 Sep 6 & 5333.32821 & 1.1 & SAAO & \\
1996 Sep 8 & 5335.31267 & 1.3 & SAAO & \\
1996 Sep 9 & 5336.37706 & 1.5 & SAAO & \\
1996 Sep 19 & 5346.36252 & 1.5 & SAAO & \\
1996 Sep 20 & 5347.34214 & 1.5 & SAAO & \\
1996 Sep 22 & 5349.34722 & 1.5 & SAAO & \\
1996 Sep 23 & 5350.37251 & 1.4 & SAAO & \\
1996 Oct 12 & 5369.28318 & 1.5 & SAAO & \\
1996 Oct 13 & 5370.26306 & 1.5 & SAAO & \\
1996 Oct 14 & 5371.27835 & 0.8 & SAAO & \\
1996 Oct 27 & 5384.27263 & 1.5 & SAAO & \\
1996 Oct 28 & 5385.26177 & 1.2 & SAAO & \\
1997 Apr 30 & 5568.55880 & 2.6 & SAAO & \\
1997 May 2 & 5570.57476 & 2.2 & SAAO & \\
1997 May 4 & 5572.59402 & 2.0 & SAAO & \\
1997 May 5 & 5573.62744 & 1.1 & SAAO & \\
1997 May 9 & 5577.59845 & 1.9 & SAAO & \\
1997 May 10 & 5578.63896 & 1.1 & SAAO & \\
1998 Apr 24 & 5927.58573 & 1.8 & SAAO & \\
1998 May 13 & 5946.57851 & 1.8 & SAAO & \\
1998 May 31 & 5964.59448 & 1.9 & SAAO & \\
1998 Jun 25 & 5989.57012 & 2.1 & SAAO & \\
1998 Jul 18 & 6012.52173 & 2.1 & SAAO & \\
1998 Jul 19 & 6014.39887 & 3.1 & SAAO & \\
1998 Jul 21 & 6016.41143 & 2.0 & SAAO & \\
1998 Aug 23 & 6049.32641 & 2.0 & SAAO & \\
1998 Aug 24 & 6050.33752 & 2.6 & SAAO & \\
1998 Sep 14 & 6071.31762 & 2.5 & SAAO & \\
1998 Sep 17 & 6074.30708 & 2.2 & SAAO & \\
1998 Sep 18 & 6075.29670 & 2.1 & SAAO & \\
1998 Oct 6 & 6093.29647 & 2.2 & SAAO & \\
1998 Oct 11 & 6098.30046 & 2.3 & SAAO & \\
1999 May 24 & 6322.60764 & 1.6 & SAAO & \\
1999 May 26 & 6324.57515 & 2.4 & SAAO & \\
1999 Jul 7 & 6367.48358 & 2.7 & SAAO & \\
1999 Jul 9 & 6368.51603 & 2.3 & SAAO & \\
1999 Aug 8 & 6399.47168 & 2.2 & SAAO & \\
1999 Aug 11 & 6402.41024 & 2.0 & SAAO & \\
1999 Aug 13 & 6404.37116 & 2.1 & SAAO & \\
1999 Aug 15 & 6406.47675 & 1.9 & SAAO & \\
1999 Aug 16 & 6407.50784 & 1.1 & SAAO & \\
1999 Sep 5 & 6427.27758 & 2.2 & SAAO & \\
1999 Sep 17 & 6439.26778 & 2.2 & SAAO & \\
1999 Oct 6 & 6458.26685 & 2.1 & SAAO & \\
1999 Oct 13 & 6465.27847 & 2.2 & SAAO & \\
2001 Jul 12 & 7103.31649 & 8.4 & SAAO & 2 \\
2001 Jul 13 & 7104.33424 & 7.9 & SAAO & 2 \\
2001 Jul 14 & 7105.30592 & 8.4 & SAAO & 2 \\
2010 Oct 14 & 10483.50415 & 4.2 & SARA-CT & 3 \\
2010 Oct 16 & 10485.55376 & 3.1 & SARA-CT & 3 \\
2011 Jun 11 & 10723.62516 & 4.0 & SARA-CT & 3 \\
2011 Jun 12 & 10724.64718 & 7.2 & SARA-CT & 3 \\
2011 Jul 20 & 10762.62790 & 4.5 & SARA-CT & 3 \\
2011 Jul 21 & 10763.53925 & 9.3 & SARA-CT & 3 \\
2011 Aug 3 & 10777.25923 & 4.3 & SAAO & 4 \\
2011 Aug 6 & 10780.26951 & 7.5 & SAAO & 4 \\
2011 Aug 7 & 10781.22805 & 8.5 & SAAO & 4 \\
2011 Aug 8 & 10782.20517 & 9.0 & SAAO & 4 \\
2011 Aug 9 & 10783.21753 & 8.6 & SAAO & 4 \\
2011 Aug 14 & 10788.22207 & 6.3 & SAAO & 4 \\
2011 Aug 15 & 10789.20804 & 8.4 & SAAO & 4 \\
2011 Aug 16 & 10790.20824 & 8.4 & SAAO & 4 \\
2011 Aug 20 & 10793.68243 & 3.8 & SARA-CT & 3 \\
2011 Aug 21 & 10794.52048 & 5.2 & SARA-CT & 3 \\
2011 Sep 19 & 10823.48896 & 5.8 & SARA-CT & 3 \\
2011 Sep 20 & 10824.52978 & 4.8 & SARA-CT & 3 \\
2011 Sep 21 & 10825.50645 & 5.3 & SARA-CT & 3 \\
2015 Jun 16 & 12189.69805 & 5.8 & SARA-CT & \\
2015 Jun 17 & 12190.71299 & 4.2 & SARA-CT & \\
2015 Jun 18 & 12191.69198 & 6.0 & SARA-CT & \\
\enddata
\tablerefs{(1)O'Donoghue et al. 1997; (2) Randall et al. 2006; (3)Otani 2015; (4)Lynas-Gray 2013.}
\end{deluxetable}

      For the 0.6-m SARA-CT observations in 2010, 2011, and 2015, a Bessel-$B$ filter was used.  The Johnson-B and Bessel-B filters lie close enough in wavelength to have negligible effect on the observed amplitudes of pulsation.  The exposure time was 40-s for all runs in this study.  For the data obtained in 2015, no filter was used, with an exposure time of 20-s because of a problem with the camera and filter wheel interface.  In order to reduce read-out noise, 2~$\times$~2 pre-binning was used for all images.  The pulsations are not expected to be wavelength independent (see \citet{Koen1998}).  However, separate analyses were performed using only the data obtained in 1995-2011 and the cumulative data obtained in 1995-2015 in order to compare the results. The O-C analysis results with and without the data obtained in 2015 were the same within the uncertainty level.  Therefore, the result including the data observed in 2015 is presented in this paper.  

\section{Data Reduction and O-C method} \label{sec:OCmethod}

      For the SARA-CT data, standard image calibration procedures were performed using the Image Reduction and Analysis Facility (IRAF)\footnote{IRAF is distributed by the National Optical Astronomy
Observatories, which is operated by the Association of Universities for
Research in Astronomy, Inc. (AURA) under cooperative agreement with the National
Science Foundation} to extract the raw intensity values for the target and comparison stars \citep{Tody1986,Tody1993}. All flat fields were exposed on a twilight sky.  For each night's data the aperture that gave the best signal-to-noise ratio (S/N) was chosen and sky annuli were used to subtract the sky background. These values were then divided by similarly extracted intensity values of non-variable comparison stars listed in Table~\ref{tab:table1}. The SAAO photomultiplier data were reduced by removing sky background as a cubic spline fitted to occasional sky measurements and then correcting for atmospheric extinction (see O'Donoghue et al. 1997 for more detail).
The SAAO CCD data were reduced using Dophot software \citep{Schechter et al.1993} which was modified by Darragh
O'Donoghue (see \citet{Randall et al.2006} and \citet{Lynas-Gray2013} for details).  For each night, the raw light curves were then normalized to the mean magnitude for that night. A second-order polynomial was used to remove mild curvature in the light curves caused by differential extinction between the target and comparison stars. All times were corrected to Barycentric Julian Date (BJD). For each night, the raw light curves were then normalized to the mean magnitude for that night.  

      The normalized light curves were then analyzed using Period04 \citep{Lenz2004}.  To improve detection and characterization of the pulsation frequencies and amplitudes, the data were pre-whitened.  The pre-whitening technique was originally introduced by \citep{Blackman1958}.  After one or more frequencies are identified in the amplitude spectrum, they are removed from each light curve by subtracting the corresponding least-squares fitted sine curve \citep{Sullivan et al.2008}.  This analysis was performed for each of the runs listed in Table~\ref{tab:table2}.  The data exhibited three pulsation peaks that matched previously published data for \objectname[V4640 Sgr]{EC 20117-4014}.  Only the two largest pulsations (F1 and F2) were 4$\sigma$ above the noise background, a threshold commonly used in the variable star community \citep [see] []{Breger et al.1999}. Therefore, only these two pulsations were used in the O-C analysis.  
      
      In the O-C method, the ``Observed" times of pulsation maxima are compared to ``Calculated" times of pulsation maxima, providing a sensitive way to detect period changes of astrophysical phenomena.  A good review of the method can be found in \citet{Paparo et al.1988}.  
      
             The O-C will be constant and flat if no period changes are occurring and the assumed period is correct.  If the calculated period is constant but incorrect, the O-C will be linear with a positive or negative slope.  If the period is changing linearly with time (e.g. due to cooling or magnetic braking), the O-C variations will exhibit a second order polynomial of form $c + bt + at^2$, where $c=\Delta E_0$ , $b=\Delta PE$, $a=\frac{1}{2}P\dot{P}E^2$ and t is time.  Here $E$ is the integer number of cycles after the first observation, $P$ is the actual period of the pulsation, $\Delta E_{0}$ is the difference between observed and calculated reference epochs, $\Delta P$ is the difference between the actual period and the estimated period (see \citet{Winget2008} for details), and $\dot{P} = dP/dt$. The precision of this technique, when applied to observations spanning several years, has allowed empirical measurement of the cooling rates of white dwarf stars and evolution of sdB stars. So far, the $\dot{P}/P$ of sdB stars, V391 Peg, HS 0702+6043, HS 0444+0458 and CS 1246 are observed \citep{Silvotti et al.2007,Lutz2011,Barlow et al.2011}.     
       
        If the O-C diagram shows periodicities, it is most likely caused by the beating of two closely spaced pulsation frequencies or reflex motion caused by an unseen companion.  The beating of two closely spaced frequencies, which may not be resolved in the power spectrum, causes not only sinusoidal variability in O-C using pulsation timings but also sinusoidal variability in pulsation $amplitudes$ \citep{Lutz2011}.  In this case, the pulsation amplitude variability is 90 degrees out of phase with pulsation timing O-C variability.  Therefore O-C sinusoidal variations due to the beating of two closely spaced frequencies are easily distinguished from the O-C periodic variations caused by a companion.  Before searching for the sinusoidal signature of a companion in the O-C diagram, a polynomial fit due to the effects of changes in pulsation period were removed.  To obtain orbital solutions from O-C diagrams, the O-C and orbital elements relationship of \citet{Irwin1952,Irwin1959} were used.  The Levenberg-Marquart (LM) algorithm \citep{Press et al.1992} was applied to evaluate the parameters of the ephemeris.
                            
        To determine ``Observed" and ``Calculated" maxima, the same method as \citet{Silvotti et al.2007}  in detecting the planet V391 Peg b was used.  At first, all pulsation periods were determined by Fourier analysis using all data sets.  Then the expected pulsation maxima (=``Calculated" maxima) were estimated by least squares fitting of a constant period sine curves using all data sets.  This equation for the largest pulsation mode (7.29 mHz) is shown below:
        
 \begin{equation}
T= 2445000.0006750 + 0.0015881 \times E 
\label{eq:1}
\end{equation}
        
On the other hand, ``Observed" maxima were determined by least squares fitting of sine curves with the same constant pulsating periods using data for $each$ night.  Here, all three detected pulsations were used for fitting the data simultaneously.  Since all pulsation periods are much shorter than one night, many pulsations occurred on a given night of observation. The time of the maximum of that fit closest to the midpoint of the night of observation was adopted as the ``Observed" time of maxima (O) estimated for that night.  Because all of a given night's data were used, this procedure maximized the precision to which a given night's phasing could be determined.  
               
      The inclination of the orbit with respect to the sky plane, $i$, is given by the following equation:
      
 \begin{equation}
a_{sdB} \sin{i} = cT
\label{eq:2}
\end{equation}
where $T$ is the  O-C amplitude, $a_{sdB}$ is the sdB star semimajor axis and $c$ is the speed of light.  Thus pulse arrival timing variations can be detected only for the pulsating sdB star, not the companion. 
The mass function for the system using O-C amplitude, $T$, is described as:

\begin{equation}
f=\frac{(M_{F}\sin{i})^3}{(M_{sdB}+M_{F})^2}=\frac{4\pi^2c^3T^3}{GP^2}
\label{eq:3}
\end{equation}
where $M_{F}$ and $M_{sdB}$ are the F  and the sdB star masses, $P$ is the orbital period, and $G$ is the gravitational constant.



 \section{Results and Discussion} \label{sec:results}
 
 \subsection{Seasonal Pulsation Amplitude Variations\label{subsec:ampshifts}}

\begin{figure}
\plotone{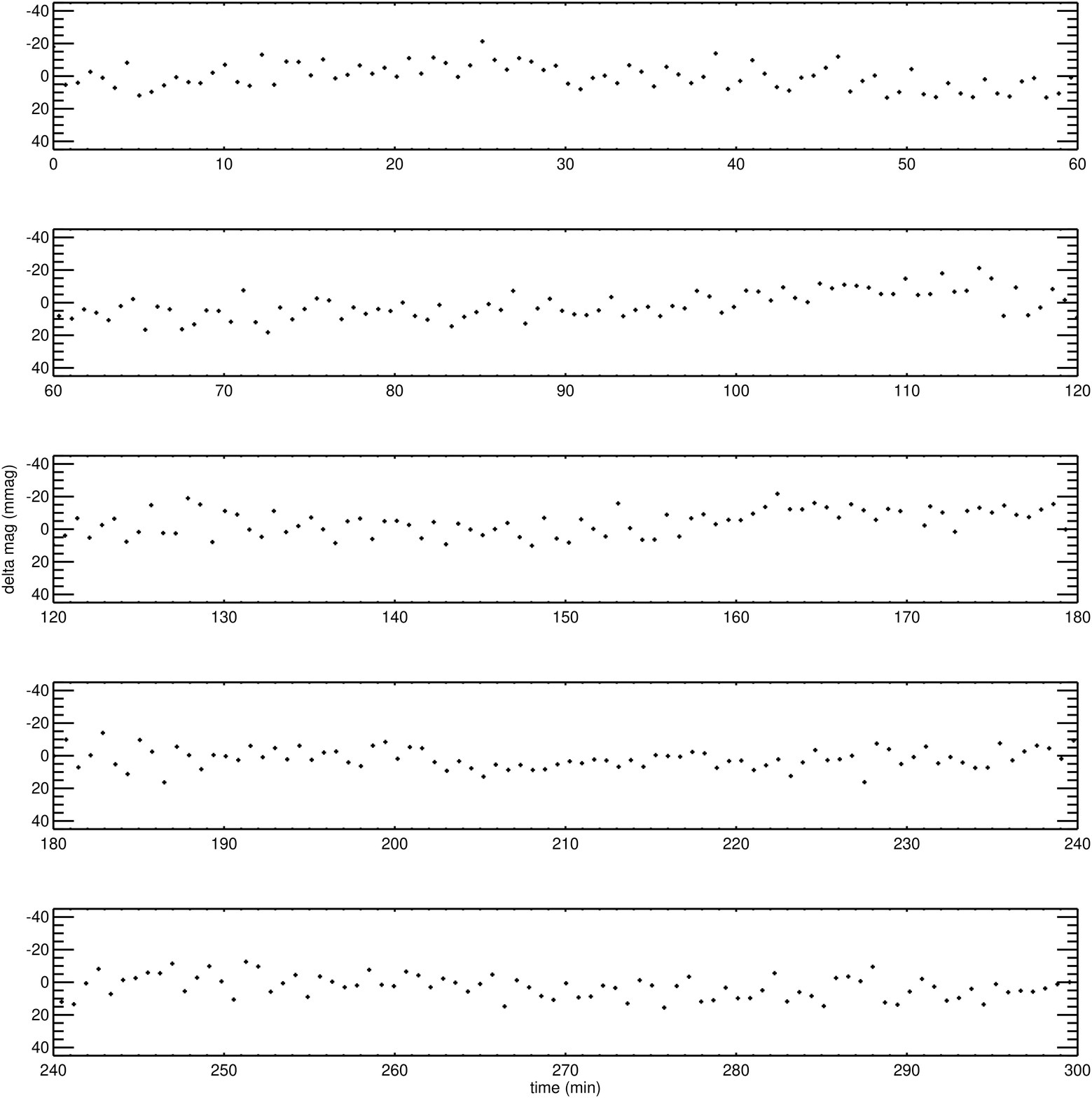}
\caption{Example light curve for EC 20117-4014 (2011 September 19).  All light curves are available in the Figure Set. \label{fig:figure2}}
\end{figure}

\begin{figure}
\plotone{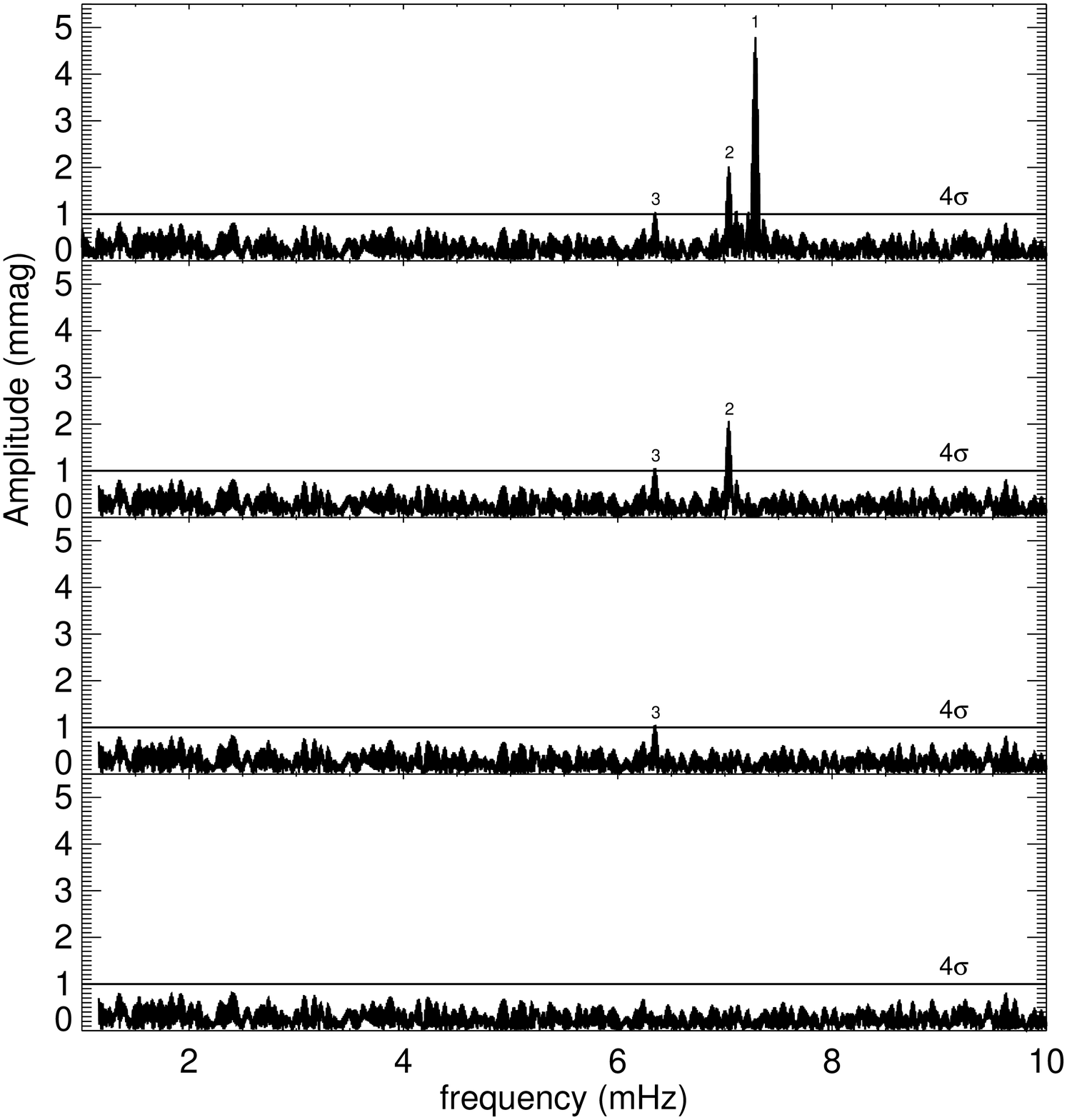}
\caption{Example Fourier analysis for EC 20117-4014 (2011 September 19-21).  The top panel shows the three most significant pulsation peaks (indicated by numbers) and the four-sigma threshold for significance adopted here. The lower panels show the successive steps of pre-whitening by sequentially removing the next largest pulsation peaks. The bottom panel shows the residual after all three pulsation peaks have been removed. In each panel, horizontal lines indicate 4$\sigma$ noise levels.  Fourier transform of each observing night data are available in the Figure Set. \label{fig:figure3}}
\end{figure}

      An example light curve for the night of 2011 September 19 and an amplitude spectrum for the 2011 September 19-21 observing run are displayed in Figures~\ref{fig:figure2} and~\ref{fig:figure3}. The highest and the lowest detectable frequencies for the data are 12.5 mHz and 0.10 mHz, respectively.  All peaks detected were between 1 mHz and 10 mHz.  Three pulsation peaks were found, which are listed in Table~\ref{tab:table3}. To obtain the frequency of the second largest pulsation, the largest pulsation was pre-whitened. To obtain the frequency of the third largest pulsation, the largest and the second largest pulsation were pre-whitened.  These peaks matched the previously published results of \citet{O'Donoghue et al.1997} and \citet{Randall et al.2006} within the formal uncertainties.  Peaks at 8.35 mHz are distinguished in some of the 1995 and 2001 data, but the S/N ratio was less than 4 $\sigma$ (this $\sigma$ is the average Fourier analysis spectrum amplitude after prewriting all three pulsations and this 8.35 mHz signal) and the peak was not detected in other data, so this peak was not included in Table~\ref{tab:table3}.  This 8.35 mHz peak is seen only in the data, which is obtained from the SAAO 0.75 m telescope.  This frequency is known to be due to the telescope drive of the SAAO 0.75-m telescope in use at the time.  

\figsetstart
\figsetnum{2}
\figsettitle{Normalized Light Curves for Each Night}

\figsetgrpstart
\figsetgrpnum{2.1}
\figsetgrptitle{May 31, 1995 to May 26, 1996}
\figsetplot{f2_1.eps}
\figsetgrpnote{Normalized light curves.}
\figsetgrpend

\figsetgrpstart
\figsetgrpnum{2.2}
\figsetgrptitle{May 27, 1996 to September 6, 1996}
\figsetplot{f2_2.eps}
\figsetgrpnote{Normalized light curves.}
\figsetgrpend

\figsetgrpstart
\figsetgrpnum{2.3}
\figsetgrptitle{September 8, 1996 to October 27, 1996}
\figsetplot{f2_3.eps}
\figsetgrpnote{Normalized light curves.}
\figsetgrpend

\figsetgrpstart
\figsetgrpnum{2.4}
\figsetgrptitle{October 28, 1996 to May 31, 1998}
\figsetplot{f2_4.eps}
\figsetgrpnote{Normalized light curves.}
\figsetgrpend

\figsetgrpstart
\figsetgrpnum{2.5}
\figsetgrptitle{June 25, 1998 to October 6, 1998}
\figsetplot{f2_5.eps}
\figsetgrpnote{Normalized light curves.}
\figsetgrpend

\figsetgrpstart
\figsetgrpnum{2.6}
\figsetgrptitle{October 11, 1998 to August 16, 1999}
\figsetplot{f2_6.eps}
\figsetgrpnote{Normalized light curves.}
\figsetgrpend

\figsetgrpstart
\figsetgrpnum{2.7}
\figsetgrptitle{September 5, 1999 to June 11, 2011}
\figsetplot{f2_7.eps}
\figsetgrpnote{Normalized light curves.}
\figsetgrpend

\figsetgrpstart
\figsetgrpnum{2.8}
\figsetgrptitle{June 12, 2011 to August 15, 2011}
\figsetplot{f2_8.eps}
\figsetgrpnote{Normalized light curves.}
\figsetgrpend

\figsetgrpstart
\figsetgrpnum{2.9}
\figsetgrptitle{August 16, 2011 to June 18, 2015}
\figsetplot{f2_9.eps}
\figsetgrpnote{Normalized light curves.}
\figsetgrpend

\figsetend

\figsetstart
\figsetnum{3}
\figsettitle{Power Spectra for Each Observation Run}

\figsetgrpstart
\figsetgrpnum{3.1}
\figsetgrptitle{May 31, 1995 to May 26, 1996}
\figsetplot{f3_1.eps}
\figsetgrpnote{Power Spectra.}
\figsetgrpend

\figsetgrpstart
\figsetgrpnum{3.2}
\figsetgrptitle{May 27, 1996 to September 6, 1996}
\figsetplot{f3_2.eps}
\figsetgrpnote{Power Spectra.}
\figsetgrpend

\figsetgrpstart
\figsetgrpnum{3.3}
\figsetgrptitle{September 8, 1996 to October 27, 1996}
\figsetplot{f3_3.eps}
\figsetgrpnote{Power Spectra.}
\figsetgrpend

\figsetgrpstart
\figsetgrpnum{3.4}
\figsetgrptitle{October 28, 1996 to May 31, 1998}
\figsetplot{f3_4.eps}
\figsetgrpnote{Power Spectra.}
\figsetgrpend

\figsetgrpstart
\figsetgrpnum{3.5}
\figsetgrptitle{June 25, 1998 to October 6, 1998}
\figsetplot{f3_5.eps}
\figsetgrpnote{Power Spectra.}
\figsetgrpend

\figsetgrpstart
\figsetgrpnum{3.6}
\figsetgrptitle{October 11, 1998 to August 16, 1999}
\figsetplot{f3_6.eps}
\figsetgrpnote{Power Spectra.}
\figsetgrpend

\figsetgrpstart
\figsetgrpnum{3.7}
\figsetgrptitle{September 5, 1999 to June 11, 2011}
\figsetplot{f3_7.eps}
\figsetgrpnote{Power Spectra.}
\figsetgrpend

\figsetgrpstart
\figsetgrpnum{3.8}
\figsetgrptitle{June 12, 2011 to August 15, 2011}
\figsetplot{f3_8.eps}
\figsetgrpnote{Power Spectra.}
\figsetgrpend

\figsetgrpstart
\figsetgrpnum{3.9}
\figsetgrptitle{August 16, 2011 to June 18, 2015}
\figsetplot{f3_9.eps}
\figsetgrpnote{Power Spectra.}
\figsetgrpend

\figsetend

\begin{deluxetable}{cccccc}
\tablecaption{Pulsation Peak Frequencies of EC 20117-4014 \label{tab:table3}}
\tablenum{3}
\tablehead{\colhead{Pulsation Mode} & \colhead{Freq} & \colhead{Freq $\sigma$ } & \colhead{Period} & \colhead{Amplitude} & \colhead{Amp $\sigma$} \\ 
\colhead{} & \colhead{(mHz)} & \colhead{(mHz)} & \colhead{(s)} & \colhead{(mmag)} & \colhead{(mmag)} } 
\startdata
F1 & 7.28484 & 1.10E-04 & 137.2 & 4.3 & 0.2 \\
F2 & 7.03529 & 2.50E-04 & 142.1 & 2.0 & 0.2 \\
F3 & 6.35145 & 5.30E-04 & 157.4 & 0.9 & 0.2 \\
\enddata
\end{deluxetable}
      
\begin{figure}
\plotone{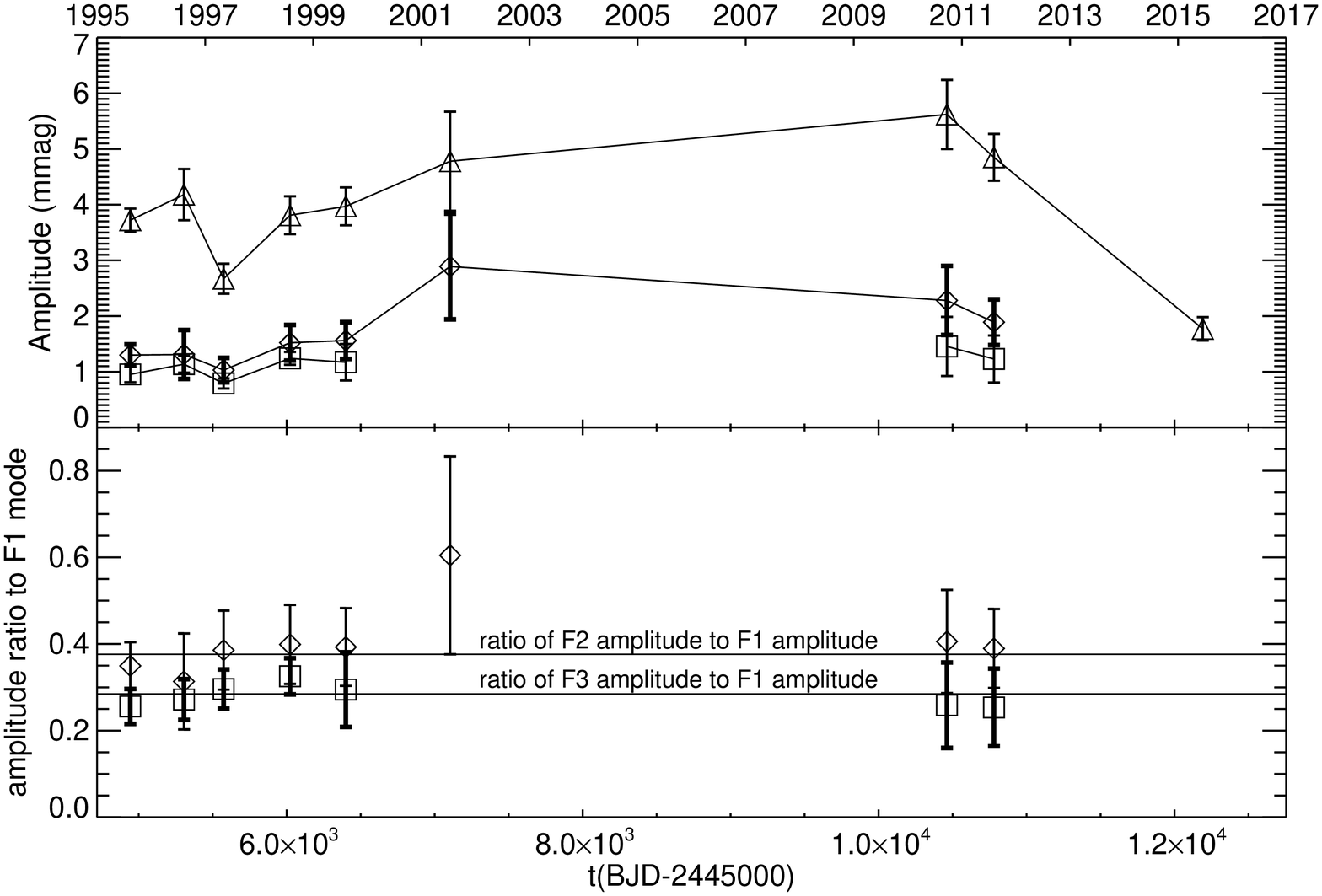}
\caption{EC 20117-4014 seasonal pulsation amplitude variations of each pulsation mode (top) and pulsation amplitude ratios of the second and third largest pulsation modes compared with the largest amplitude pulsation mode of 7.29mHz (bottom).  $Top:$ Data points of the largest amplitude pulsation mode F1: 7.29 mHz are marked by triangles.  The points of the second and third largest amplitude pulsation modes, F2 (7.04 mHz) and F3 (6.35 mHz), are marked by diamonds and squares, respectively. Each data point is the weighted average amplitude during each year (1995, 1996, 1997, 1999, 2001, 2010, 2011, and 2015). The F2 mode was not detected in the data observed in 2015, and the F3 mode was not detected in the data observed in 2001 and 2015.  The F1 mode and F2 modes show obvious pulsation amplitude variations. Seasonal amplitude variations appear to be consistent for all three frequencies, although F3 mode amplitude variations are too small (compared to the uncertainty) to confirm the variations are real.  $Bottom:$ F2 mode (diamonds) and F3 mode  (squares) pulsation amplitudes compared to F1 amplitudes.  Both ratios fit straight lines.  The horizontal lines are the weighted average ratios of both frequencies, which is 0.38 $\pm$ 0.03 for amplitude ratios F2/F1, and 0.28 $\pm$ 0.02 for F3/F2. \label{fig:figure4}}
\end{figure}

      Since seasonal amplitude variations were detected in the largest pulsation frequency by \citet{Otani2015}, the possibility of seasonal amplitude changes was investigated. Seasonally-binned pulsation amplitude variations as a function of date are shown in the top panel of Figure~\ref{fig:figure4}.  These seasonal pulsation amplitude data points are weighted mean amplitudes of each year's data. This weight was obtained from $w = 1/\sigma_{i}^{2}$, where $\sigma_{i}$ is the pulsation amplitude uncertainty for each night. Triangles represent amplitudes of the largest amplitude pulsation frequency (F1: 7.29 mHz). Diamonds represent amplitudes of the second largest amplitude pulsation frequency (F2: 7.04 mHz), and squares denote amplitudes of the third largest amplitude pulsation frequency (F3: 6.35 mHz).  The F2 mode was not detected in the data observed in 2015, and the F3 mode was not detected in the data observed in 2001 and 2015. These seasonal amplitude changes appear to be consistent for all three frequencies, although the F3 variations are comparable with the uncertainties and their reality is questionable. This figure shows that the amplitude of the F1 mode has been decreasing since 2010 at a rate of 0.0021 mmag per day.  The pulsation amplitude of the data observed in 2015 is almost one third of the pulsation amplitude of the data observed in 2010.  To test whether the changes in amplitudes are correlated, we compared the ratio of the seasonal amplitude of F1 to F2 and to F3.  These results are shown in the bottom panel of Figure~\ref{fig:figure4}.  Both ratios fit zero-slope straight lines within their uncertainties.  The weighted average ratios are 0.38 $\pm$ 0.03 for F2/F1, and 0.28 $\pm$ 0.02 for F3/F1.  While the seasonal amplitude variation is consistent in all pulsations, the F1
amplitude decline from 2010 to 2015 (in particular) cannot be understood to
be a consequence of rotational splitting.  Nonetheless it is of interest to note
that \citet[their figure 1]{Hutchens et al.2017} observe a similar amplitude decline
in the single-mode sdB pulsator CS 1246.

    \subsection{O-C Periodic variation due to the Existence of the Companion\label{subsec:OCvariation}}
 
The first and second largest amplitude pulsations F1 and F2 were used for an O-C analysis. The S/N ratios of the third largest pulsations (F3) for each night were less than 4 $\sigma$, so this mode was not used in our analysis.  Two distinct causes of period variation were identified in both F1 and F2 (O-C) values.  In addition to the near sinusoidal variation due to light travel-time changes caused by the reflex motion of the pulsating sdB star, a second order polynomial variation attributable to sdB star evolution was also discernible.  Orbital and evolution contributions were fitted simultaneously; in fitting orbits,
a single offset was established so as to minimize the difference between the F1 
and F2 orbits.  The pulsation period change is $\dot{P}$ = 5.4 ($\pm$ 0.7) $\times$ 10$^{-14}$ d d$^{-1}$.  The removed second order polynomial curves for O-C diagrams is:

\begin{equation}
O-C = (2.6 (\pm 0.4) \times 10^{-6}) t^{2} - (4.0 (\pm 0.6) \times 10^{-2}) t + 1.5 (\pm 0.2) \times 10^{2}
\label{eq:4}
\end{equation}
Here, t is BJD - 2445000.


The rate of period change ($\dot{P}$) indicates the age of the sdB star after the zero-age extreme horizontal branch (ZAEHB) \citep{Charpinet et al.2002}.  For p-modes,  $\dot{P}$ is positive during the first evolutionary phase, which is before the thermonuclear fuel in its center is exhausted.  $\dot{P}$ is negative during the second evolutionary phase, which is after the depletion of thermonuclear fuel in its center and before the post-EHB evolution.  The change of sign occurs around 87-91 Myr after the ZAEHB. The positive values of $\dot{P}$ for \objectname[V4640 Sgr]{EC 20117-4014} (sdB) denote that the the star is still in the first evolutionary phase.  The age of \objectname[V4640 Sgr]{EC 20117-4014} (sdB) can also be estimated from its effective temperature  and surface gravity~$g$. Figure~1 of \citet{Fontaine et al.2012} indicates that the age of EC 20117-4014 (sdB) is 90 $\pm$ 5 Myr.  Therefore the sdB component of \objectname[V4640 Sgr]{EC 20117-4014}  is about to end its core helium-burning phase. 

Also, we can set an upper limit on the age of the sdB star since the zero-age main sequence (ZAMS).  \citet{Choi et al.2016} suggests that the Main Sequence (MS) lifetime of a 1.5 $M_{\odot}$ star, like the late F main-sequence companion, is about 2.5 Gyr (See Figure~12 of their article).  Assuming that both the sdB star and its F companion are coeval, an upper limit to the age of the sdB star is therefore 2.5 Gyr.  However, this argument is not valid if large mass exchange from sdB progenitor to the companion occured during the RLOF evolution.

The time scale for radius change is also obtained from the time scale for period change: 

\begin{equation}
\frac{\dot{P}}{P} \approx \frac{3}{2}  \frac{\dot{R}}{R}
\label{eq:6}
\end{equation}
Here, $R$ is the radius of the star.  For \objectname[V4640 Sgr]{EC 20117-4014}, the time scale for period change ($P / \dot{P}$) is roughly $ 8.0 \times 10^{7}$ yr.  This value corresponds to a time scale for the radius increase of $R / \dot{R} \approx 1.2 \times 10^{8}$ yr.  According to Figure~3 of \citep{Kawaler2010}, $R / \dot{R} \approx 2 \times10^{8}$ yr at the first phase of core helium burning and it will turn negative at the second phase of shell burning. $R / \dot{R}$ for the sdB component of \objectname[V4640 Sgr]{EC 20117-4014} is about a half of $2 \times10^{8}$ yr, and this also indicates that this star's core-burning is about to terminate. 

\begin{figure}
\plotone{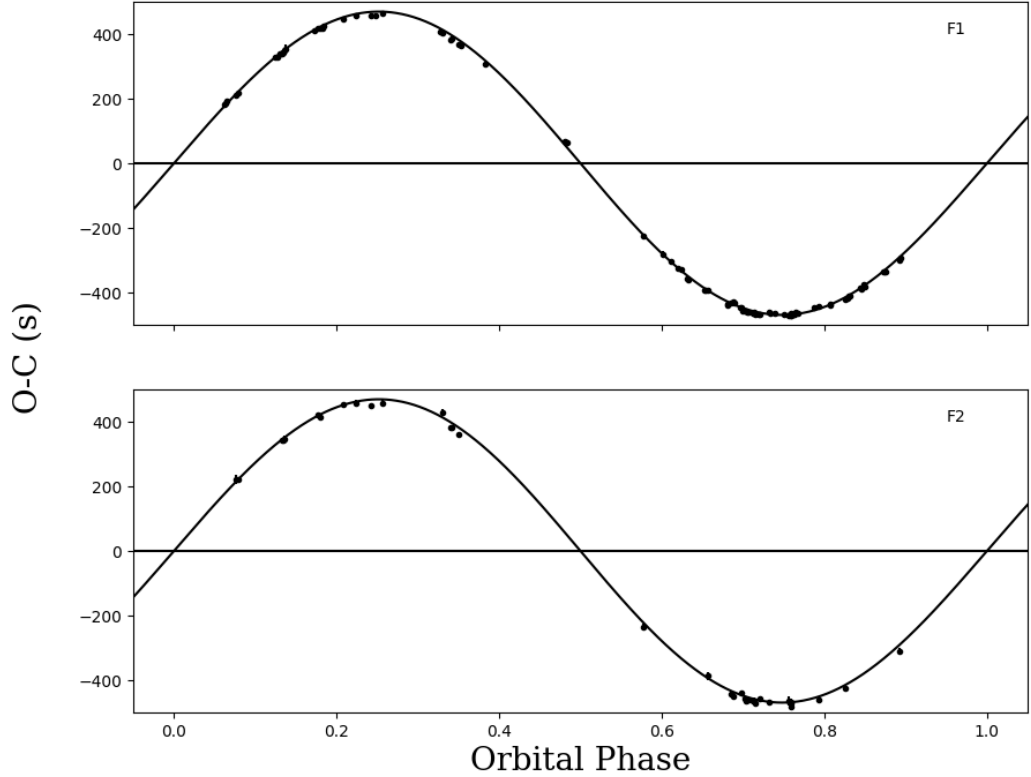} 
\caption{Phase-folded O-C curves for EC 20117-4014 constructed from the largest and the second largest amplitude pulsations (F1: 7.29 mHz at the top and F2: 7.04 mHz at the bottom). The third largest amplitude pulsation of each night was lower than 4 $\sigma$ above the noise level so it was not used for the O-C analysis.  Each square represents the O-C value of each entire night.  O-C uncertainties of some data points are smaller than the symbol size. Periods and amplitudes of fitted curve is 792.3 d and 468.9 s. \label{fig:figure5}} 
\end{figure}

Figure~\ref{fig:figure5} presents the phase-folded O-C diagrams for F1 and F2 after the removal of polynomials given by equation (\ref{eq:4}).  For F2, the only nights shown are those for which the pulsation
amplitude is 4-$\sigma$ above the noise level. The pulsation periods used are 137.2729-s for F1 and 142.1477-s for F2.  Table~\ref{tab:table5} lists all O-C data points for F1 and F2 pulsation modes respectively.  The solid curves in Figure~\ref{fig:figure5} are the best fitting orbital solutions. The orbital periods are 792.3 $\pm$ 0.3 d, and these are the same within the uncertainties.  Initially, we inadvertently omitted 1998 data from the (O-C) diagram.  When the 1998 data were included, these fitted the already established O-C curve well.  The best fitted orbital solutions for F1 and F2 O-C data points are shown in Table~\ref{tab:table6}. The formal $\chi ^2$ values are  86.5 (F1) and 27.4 (F2), respectively.  Degrees of freedom of F1 and F2 are 84 and 30.  The corresponding right tail p-values are 0.40 and 0.60.  Therefore, the correlation between the periodicity of the O-C data points and the binary motion with P=792.3 days are not rejected. The chi-squared values being consistent with the number of degrees of freedom
suggest that all relevant physical information has been extracted from the data;
in particular, there is no evidence for a third body associated with the EC 20117-4014
binary which would not have been accounted for in our model.

\startlongtable
\begin{deluxetable}{ccccc}
\tablecaption{O-C data points for F1 (7.29 mHz) and F2 (7.04 mHz) pulsation modes before removing the second order polynomial fit \label{tab:table5}}
\tablenum{4}
\tablehead{\colhead{Time} & \colhead{[F1]O-C} & \colhead{[F1] O-C uncertainties} & \colhead{[F2]O-C} & \colhead{[F2] O-C uncertainties} \\ 
\colhead{(BJD-2445000)} & \colhead{(s)} & \colhead{(s)} & \colhead{(s)} & \colhead{(s)} } 

\startdata
4869.492986 & 470.0 & 2.0 & 472.0 & 5.5 \\
4947.402781 & 396.7 & 1.9 & 402.7 & 4.9 \\
4948.291201 & 397.5 & 2.9 & 402.0 & 8.9 \\
4955.328631 & 381.2 & 2.7 & 381.8 & 6.7 \\
4957.252794 & 376.9 & 2.3 &  &  \\
4980.266303 & 319.2 & 4.6 &  &  \\
5168.588619 & -315.1 & 3.6 &  &  \\
5194.611691 & -383.6 & 5.6 &  &  \\
5195.609237 & -381.7 & 4.6 &  &  \\
5229.582828 & -438.1 & 5.2 &  &  \\
5230.582731 & -435.7 & 4.5 & -426.2 & 10.6 \\
5262.537515 & -457.3 & 7.0 &  &  \\
5272.487817 & -460.3 & 6.9 &  &  \\
5276.466287 & -462.0 & 5.6 & -451.1 & 18.5 \\
5281.489239 & -457.9 & 5.5 &  &  \\
5285.456137 & -454.3 & 4.8 &  &  \\
5316.471197 & -432.4 & 5.0 &  &  \\
5317.494223 & -427.2 & 6.3 &  &  \\
5331.450016 & -412.3 & 5.6 & -415.5 & 14.9 \\
5333.350491 & -409.7 & 8.1 &  &  \\
5335.340110 & -404.4 & 4.5 &  &  \\
5336.408112 & -402.8 & 7.3 &  &  \\
5346.393672 & -378.5 & 5.7 &  &  \\
5347.373323 & -383.0 & 5.8 &  &  \\
5349.378340 & -366.6 & 7.1 &  &  \\
5350.400659 & -375.8 & 7.6 &  &  \\
5369.315228 & -329.4 & 4.4 &  &  \\
5370.294472 & -326.8 & 6.3 &  &  \\
5371.294833 & -330.3 & 5.4 &  &  \\
5384.303378 & -291.9 & 8.0 & -302.2 & 11.7 \\
5385.286270 & -287.2 & 4.6 &  &  \\
5568.558797 & 332.6 & 3.1 &  &  \\
5570.574757 & 334.5 & 4.7 &  &  \\
5572.594017 & 343.5 & 7.5 &  &  \\
5573.627437 & 342.5 & 5.6 &  &  \\
5576.640367 & 354.5 & 3.5 &  &  \\
5577.598447 & 354.2 & 3.7 & 351.8 & 12.1 \\
5578.638957 & 358.3 & 14.2 &  &  \\
5927.623752 & -223.2 & 4.3 & -233.5 & 13.4 \\
5946.615308 & -282.1 & 11.0 &  &  \\
5954.636558 & -302.7 & 3.7 &  &  \\
5964.633991 & -326.6 & 4.0 &  &  \\
5989.612737 & -391.4 & 4.1 & -385.8 & 12.4 \\
6012.564885 & -430.5 & 4.2 & -443.5 & 8.4 \\
6014.466184 & -429.7 & 3.6 & -449.9 & 9.1 \\
6016.453894 & -433.5 & 4.2 &  &  \\
6049.365569 & -462.3 & 6.4 & -466.5 & 11.5 \\
6050.390923 & -464.1 & 3.0 &  &  \\
6071.370080 & -464.4 & 3.1 & -466.8 & 7.9 \\
6074.351007 & -466.9 & 3.9 &  &  \\
6075.339904 & -462.8 & 3.9 &  &  \\
6093.341990 & -447.1 & 6.4 &  &  \\
6098.348228 & -444.6 & 3.5 & -459.3 & 9.6 \\
6322.602836 & 206.6 & 4.7 & 219.0 & 13.3 \\
6324.570313 & 214.2 & 3.5 & 218.9 & 6.0 \\
6367.457465 & 337.4 & 3.6 & 342.1 & 6.9 \\
6368.507951 & 341.1 & 3.1 & 340.2 & 10.1 \\
6399.464757 & 407.8 & 4.3 &  &  \\
6402.403646 & 415.7 & 4.0 & 420.2 & 9.3 \\
6404.364641 & 415.0 & 3.6 & 412.1 & 8.1 \\
6406.470197 & 413.9 & 4.4 &  &  \\
6407.501447 & 420.4 & 7.3 &  &  \\
6427.272280 & 443.8 & 5.0 & 450.8 & 9.9 \\
6439.263252 & 453.4 & 3.8 & 456.6 & 10.8 \\
6458.263947 & 454.5 & 4.1 &  &  \\
6465.276215 & 459.9 & 3.2 & 456.4 & 10.6 \\
7103.488750 & 173.9 & 9.4 &  &  \\
7104.488620 & 177.6 & 6.1 &  &  \\
7105.474873 & 186.2 & 6.3 &  &  \\
10483.589470 & 417.7 & 5.7 &  &  \\
10485.617570 & 415.3 & 6.0 & 423.4 & 13.2 \\
10723.804710 & -342.7 & 7.0 &  &  \\
10724.797720 & -347.6 & 4.8 &  &  \\
10762.721360 & -420.4 & 7.6 &  &  \\
10763.733280 & -423.9 & 4.6 &  &  \\
10777.350960 & -441.2 & 3.7 &  &  \\
10780.428190 & -443.8 & 2.5 & -440.3 & 5.4 \\
10781.391280 & -444.0 & 4.1 & -450.6 & 8.4 \\
10782.395440 & -445.4 & 3.0 & -444.7 & 8.5 \\
10783.396050 & -444.4 & 2.9 & -445.6 & 9.0 \\
10788.339390 & -448.9 & 3.3 & -448.5 & 16.1 \\
10789.383570 & -447.5 & 3.0 & -447.5 & 8.2 \\
10790.383670 & -451.5 & 2.3 & -455.8 & 6.6 \\
10793.761090 & -451.5 & 5.6 &  &  \\
10794.629430 & -452.3 & 3.6 & -442.2 & 8.8 \\
10823.610400 & -454.1 & 4.0 & -450.6 & 8.2 \\
10824.618790 & -454.9 & 4.4 & -454.2 & 9.2 \\
10825.615890 & -456.4 & 3.6 & -464.2 & 11.4 \\
12189.819940 & 108.8 & 4.8 &  &  \\
12190.800550 & 106.1 & 5.8 &  &  \\
12191.817640 & 104.6 & 5.4 &  &  \\
\enddata
    \tablecomments{Time is mid-observing time.}
\end{deluxetable}

    The resulting period of a periodic curve (P= 792.3 d) does not match sinusoidal variabilities in the F1 and F2 pulsation amplitudes, so these are not due to the beating of two closely spaced pulsation frequencies.  Therefore, we concluded that the resulting variations are due to the light-travel effects caused by the F-type companion. 
   
   Most subdwarf-B (sdB) stars in binary systems have companions which
are white dwarfs or M-dwarf Main Sequence stars
\citep{2015A&A...576A..44K}; these have short 
orbital periods $(\lesssim 10\ {\rm days})$ and are understood 
\citep{Han et al.2002,Han et al.2003,2017A&A...599A..54X} 
to be a consequence of evolution in a common envelope.
A few sdB binaries have longer orbital periods with a F- or G-type giant or
Main Sequence star; ${\rm EC}~20117-4014$ was confirmed in the
present paper to be an example and
\citet{AznarCuadrado2002}, \citet{2012MNRAS.427.2180N} and \citet{Vos2017} discovered some additional
systems.
SdB stars in wide binaries are formed 
\citep{Han et al.2002,Han et al.2003} 
as a consequence of a
red giant progenitor losing almost all of its hydrogen envelope,
at the onset of core helium-burning, through stable Roche lobe
overflow (RLOF);
their calculations suggest that the orbits should be circular and
have periods $\lesssim 500\ {\rm days}$.

Orbital element determination for long-period sdB binaries presents a
greater challenge than for those with short periods.
Radial velocity observations by \citet{2012ASPC..452..163O, Deca et al.2012, Barlow et al.2013, Wade et al.2014}, identified sdB stars having a main sequence or giant companion with orbital periods $> 500\ {\rm days}$, significantly greater
than the \citet{Han et al.2002,Han et al.2003} orbital-period
distribution would suggest.
\citet{2013MNRAS.434..186C} reproduce the orbital-period distribution
\citet{2012ASPC..452..163O} observe using detailed binary evolution 
calculations for the stable RLOF channel, improving on the simplified
binary population synthesis by \citet{Han et al.2003}. The estimated period of binaries that went through this RLOF channel is P = 400 - 1100 d.
The derived orbital period of the EC 20117-4014 binary system (P = 792.3 d) falls in the middle of this range. 
   
\begin{figure}
\plotone{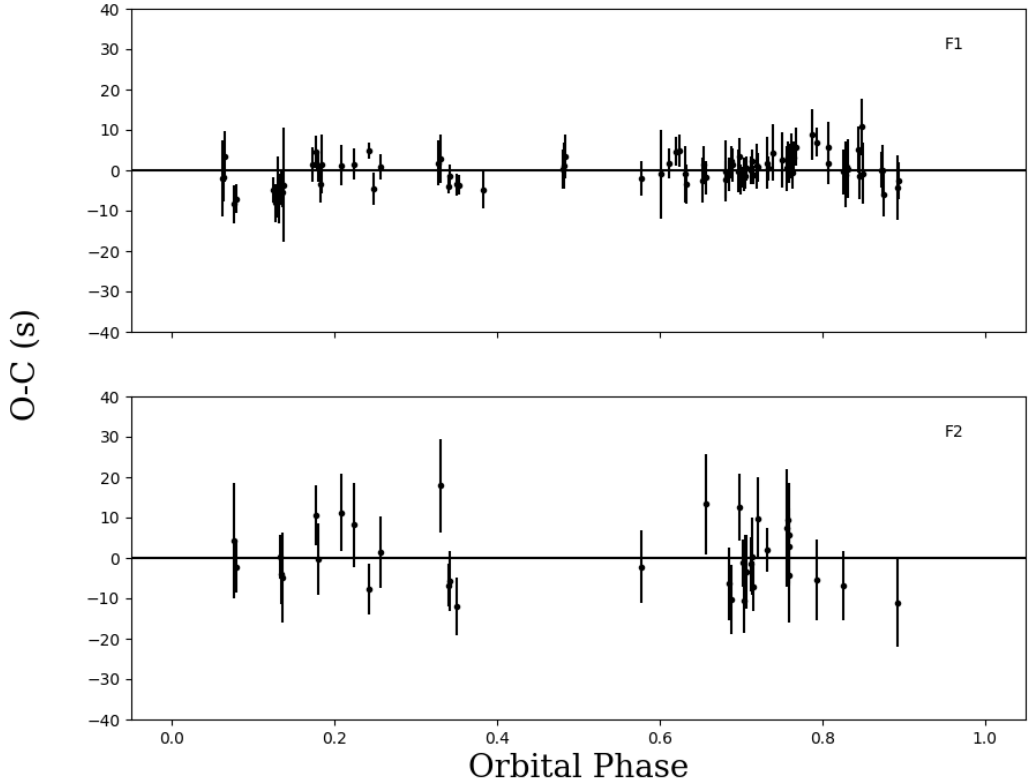}
\caption{Residuals of the fitting of O-C data points shown in Figure~\ref{fig:figure5}. The top panel show the residual for F1 O-C data points and the bottom panel show the residual for F2 O-C data points. \label{fig:figure6}}
\end{figure}

   Figure~\ref{fig:figure6} shows the residuals of the fitted orbital solution of O-C data points shown in Figure~\ref{fig:figure5}.  
   \citet[their figure 1]{2015A&A...579A..49V} plot orbital eccentricities 
against periods for those long-period sdB binaries for which orbital
elements had been determined at the time of publication.  A clear
correlation is apparent with longer period systems having higher orbital
eccentricities.  Modifications to the binary module of the stellar
evolution code Modules for Experiments in Stellar Astrophysics
\citep[MESA]{2011ApJS..192....3P,2013ApJS..208....4P,2015ApJS..220...15P}
by \citet{2015A&A...579A..49V}, to include eccentricity pumping 
processes, results in binary systems with observed orbital
eccentricities when eccentricity pumping via a circumbinary disk is
accompanied by phase-dependent RLOF.  A remaining difficulty is the 
model prediction of some high orbital eccentricities for short periods
which are at variance with available observations.

The deduced period of 792.3 days
suggested an orbital eccentricity of 0 - 0.07 on the basis of the
\citet{2015A&A...579A..49V} figure~1.  No significant eccentricity is detected in the EC 20117-4014 binary system, and the upper limit to the eccentricity is 0.025 if 3$\sigma$ is used as an upper limit.  The eccentricity of this binary system has the lowest upper limit among all wide sdB binaries with known orbital parameters \citep{Vos2017}.
       
   Using the periodicity shown in Figure~\ref{fig:figure5} (792.3 $\pm$ 0.3 d), which is due to the orbital motion, the semi-major axis for the star's orbit was calculated. If both the sdB star mass ($m_{sdB}$) and F star mass($m_{F}$) are known, semi-major axes of both sdB and F main sequence stars are calculated using Equations (\ref{eq:2}) and (\ref{eq:3}). According to \citet{Randall et al.2006}, the EC20117-4014 sdB star mass range is $m_{sdB} = 0.50 - 0.59 M_{\odot}$.  However, the F star mass of the EC 20117-4014 system is unknown.  Therefore, the F stars' mass was estimated from the stellar type.  Since the estimated stellar type of this companion is F5V or late F main-sequence \citep{O'Donoghue et al.1997}, the main-sequence mass range \citep{Schmidt-Kaler1982} of F5-G0 (which is 1.40-1.05 $M_{\odot}$) was used to estimate the semi-major axis.  The semi-major axes ranges of the sdB star and the F main sequence star are $a_{sdB}$ = 1.3-1.5 AU and $a_{F}$ = 0.5-0.7 AU.   
   The inclination was computed using possible mass ranges.  The inclination range obtained is $i$ = 37.9\degr - 48.0\degr. It is not surprising that eclipses have not been detected in the observed light curves. The orbital parameters are shown in Table~\ref{tab:table6}.
   




\begin{deluxetable}{lc}
\tablecaption{Orbital information of the EC 20117-4014 system \label{tab:table6}}
\tablenum{5}
\tablehead{\colhead{Parameters} & \colhead{values} \\ 
\colhead{} & \colhead{} } 
\startdata
Period (days) & 792.3 $\pm$ 0.3 \\
Amplitude (s) & 468.9 $\pm$ 1.1 \\
eccentricity & 0.004 $\pm$ $^{0.007}_{0.004}$ \\
zero point of time (BJD-2445000) & 4673.0 $\pm$ 0.4 \\
radial velocity for sdB star $K_{sdB}$ (km/s) & 12.90 $\pm$ 0.03 \\
semimajor axis $a_{sdB}$ (AU) & 1.3 - 1.5 \\
semimajor axis $a_{F5}$ (AU) & 0.5 - 0.7 \\
inclination (degrees) & 37.9 - 48.0 \\
\enddata
\tablecomments{semimajor axes $a_{F}\& a_{sdB}$ and inclination $i$ are estimated ranges using possible sdB and F main sequence stars masses.}
\end{deluxetable}
   
      In principle, the F main sequence companion should be detectable via radial velocity variations.  However, radial velocities are difficult to measure for sdB stars because of their high gravity (which broadens the line profiles). Generally, the (non-relativistic) radial velocity of the sdB star can be estimated using the mass function: 
      
\begin{equation}
f=\frac{(M_{F}\sin{i})^3}{(M_{sdB}+M_{F})^2}=\frac{PK^3}{2\pi G}
\label{eq:14}
\end{equation}
where $M_{sdB}$ and $M_{F}$ are the masses of the sdB star and F companion.  $P$, $K$, and $G$ are the orbital period, radial velocity semi-amplitude, and gravitational constant, respectively. Comparing this mass function with the previously obtained mass function using O-C amplitude, $T$, (see equation (\ref{eq:2})), the relationship between $K$ and $T$ is described as:

\begin{equation}
K=\frac{2Tc\pi}{P}
\label{eq:15}
\end{equation}

This equation only depends on $T$ and $P$, and is not dependent on $M_{sdB}$ and $M_{F}$.  Using this equation yields an estimate for the sdB star radial velocity amplitude of 12.90 $\pm$ 0.03 km s$^{-1}$. 
 
 \citet{Silvotti et al.2014} succeeded in measuring the radial velocities of sdB stars to a precision of 50-100 m s$^{-1}$ (5-sigma level) using the 3.6 m $Telescopio\ Nazionale\ Galileo$ (TNG).  The EC 20117-4014 sdB star radial velocity amplitude obtained is much larger than this, so it is possible to confirm the companion using the radial velocity method. However, as shown here, much smaller telescopes (1-m class) are adequate to detect it using the O-C method.
 
  \section{Conclusions}\label{sec:conc}

This paper presents an orbital solution for the EC 20117-4014 binary and the evolutionary stage of its sdB component.  The formation theory for sdB stars is still hotly debated and the frequency with which companions are found may help settle this issue.  In particular, orbital element determination for long-period sdB + F/G dwarf binaries is not as common as those with short periods, and only 12 binaries' orbital solutions were obtained so far.  Among those systems, ${\rm EC}\:20117\!-\!4014$ is unique in that the orbital elements were 
obtained using the changing light travel time across the sdB star
orbit, not using radial velocities.  Photometric data of EC20117-4014 for 20 years shows obvious periodic variations in both the largest and the second largest amplitude pulsation frequencies and allowed an orbital solution for the F main-sequence companion and the sdB star.  The data indicate the orbital parameters of the binary system are:\\

\begin{enumerate}
\item Orbital period P = 792.3 $\pm$ 0.2  days
\item The upper limit of the eccentricity is 0.025. This is the lowest upper limit of the eccentricity among all 12 wide sdB binaries with known orbital parameters.  
\item The light-travel time amplitude A = 468.9 $\pm$ 1.1 s
\item Inclination $i = 37.9\degr - 48.0 \degr,$ which does not permit eclipses.
\end{enumerate}

Importantly, the \objectname[V4640 Sgr]{EC 20117-4014} sdB star represents another example of a system that has very likely experienced binary evolution, and lost almost all of its hydrogen envelope through stable RLOF.  The 20 years of photometric data also showed the time scale for period changes $\dot{P} / P$.  The $\dot{P} / P$ and the corresponding time scale for radius increase indicates that the \objectname[V4640 Sgr]{EC 20117-4014} is about to deplete thermonuclear fuel in its center. 

\acknowledgments{Acknowledgments}

TDO gratefully acknowledges support from National Science Foundation grants AST-108845, AST-1358787, and PHY-1358879.\\

AELG is indebted to the panel for the Allocation of Telescope Time (United Kingdom) for the award of travel and subsistence expenses.\\

This paper uses observations made at the South African Astronomical Observatory and we grateful for generous allocations of telescope time 
by the SAAO and for the continuing support at the telescope provided by SAAO staff. This paper is based partly upon work supported financially by 
the National Research Foundation of South Africa.\\

We are indebted to an anonymous referee for comments and suggestions which have improved our paper.\\

We are indebted to Dr. J. Vos for providing us with a copy of his elliptical orbit fitting program.

%

\vspace{5mm}
\facilities{SAAO: 0.5m, SAAO: 0.75m, SAAO: 1.0m, CTIO:0.6 m}


\software{Period04 \citep{Lenz2004}}


\begin{thebibliography}{}

\bibitem[Aznar Cuadrado \& Jeffery(2002)]{AznarCuadrado2002}
Aznar Cuadrado, R. \& Jeffery, C. S., 2002, \aas, 385, 131

\bibitem[{{Applegate}(1992)}]{1992ApJ...385..621A}
{Applegate}, J.~H. 1992, \apj, 385, 621

\bibitem[Barlow et al.(2011)]{Barlow et al.2011}
Barlow, B. N., Dunlap, B. H., Clemens, J. C., Reichart, D. E., Ivarsen, K. M., LaCluyze, A. P., Haislip, J. B., \& Nysewander, M. C., 2003, \aap, 398, 283
 
\bibitem[Barlow et al.(2013)]{Barlow et al.2013}
Barlow, B. N., Liss, S. E., Wade, R. A., \& Green, E. M., 2012, \apj, 771, 23
 
\bibitem[Baran et al.(2012)]{Baran et al.2012}
Baran, A. S., Reed, M. D., Stello, D., et al. 2012, \mnras, 424, 2686

\bibitem[Bonanno et al.(2003)]{Bonanno et al.2003}
Bonanno, A., Catalano, S., Frasca, A., Mignemi, G., \& Patern\`{o}, L., 2003, \aap, 398, 283

\bibitem[Bergeron et al.(1994)]{Bergeron et al.1994}
Bergeron, P., Wesemael, F., Beauchamp, A., Wood, M. A., Lamontagne, R., Fontaine, G., \& Liebert, James, 1994, \apj, 432, 305 

\bibitem[Breger et al.(1999)]{Breger et al.1999}
Breger M. et al., 1999, \aap, 349, 225

\bibitem[Blackman, \& Tukey(1958)]{Blackman1958}
Blackman, R. B. \& Tukey, J. W., 1958, Bell Syst. Tech. J., 398, 283

\bibitem[Charpinet et al.(1996)]{Charpinet et al.1996}
Charpinet, S., Fontaint, G., Brassard, P., \& Dorman, B., 1996, \apjl, 471, L103

\bibitem[Charpinet et al.(2000)]{Charpinet et al.2000}
Charpinet, S., Fontaine, G., Brassard, P., \& Dorman, B., 2000, \apjs, 131, 223

\bibitem[Charpinet et al.(2002)]{Charpinet et al.2002}
Charpinet, S., Fontaint, G., Brassard, P., \& Dorman, B., 2002, \apss, 140, 469

\bibitem[{{Chen} {et~al.}(2013){Chen}, {Han}, {Deca}, \&
  {Podsiadlowski}}]{2013MNRAS.434..186C}
{Chen}, X., {Han}, Z., {Deca}, J., \& {Podsiadlowski}, P. 2013, \mnras, 434,186

\bibitem[{{Chen} \& {Podsiadlowski}(2017)}]{2017ApJ...837L..19C}
{Chen}, W.-C., \& {Podsiadlowski}, Ph. 2017, \apjl, 837, L19

\bibitem[Choi et al.(2016)]{Choi et al.2016}
Choi, J., Dotter, A., Conroy, C., Cantiello, M., Paxton, B.,  \& Johnson, B. D., 2016, \apj, 823, 2

\bibitem[Clausen \& Wade (2011)]{Clausen et al.2011}
Clausen, D., \& Wade, R. A., 2011, \apj, 733, L42

\bibitem[Deca et al.(2012)]{Deca et al.2012}
Deca, J., Marsh, T. R., \& {\O}stensen, R. H. et al., \mnras, 421, 2798

\bibitem[Dorman, Rood \& O'Connell(1993)]{Dorman et al.1993}
Dorman, B., Rood, R. T., \& O'Connell, R. W., 1993, \apj, 419, 596

\bibitem[Efron(1979)]{Efron1979}
Efron, B., 1979, Ann. Statist, 7, 1, 1-26

\bibitem[Fontaine et al.(2012)]{Fontaine et al.2012}
Fontaine, G., Brassard, P., Charpinet, S., Green, E. M., Randall, S. K., \& Van Grootel, V., 2012, \aap, 539, 12

\bibitem[Green et al.(2003)]{Green et al.2003}
Green, E. M. et al., 2003, \apj, 583, 31

\bibitem[Han et al.(2002)]{Han et al.2002}
Han, Z., Podsiadlowski, Ph., Maxted, P. F. L., Marsh, T. R., \& Ivanova, N., 2002, \mnras, 336, 449

\bibitem[Han et al.(2003)]{Han et al.2003}
Han, Z., Podsiadlowski, Ph., Maxted, P. F. L., \& Marsh, T. R., 2003, \mnras, 341, 669

\bibitem[Heber(2009)]{Heber2009}
Heber, U., 2009, \araa, 47, 211

\bibitem[Hinkley(1988)]{Hinkley1988}
Hinkley, D., 1988, J. Royal Stat. Soc. Series B, 50, 3, 321

\bibitem[Hutchens et al.(2017)]{Hutchens et al.2017}
Hutchens, Z. L., Barlow, B. N., Vasquez, S. A., Reichart, D. E., Haislip, J. B., Kouprianov V. V., Linder, T. R. \& Moore, J. P., 2017, Open Astronomy, 26, 252

\bibitem[Irwin(1952)]{Irwin1952}
Irwin, J. B., 1952, \apj, 116, 211

\bibitem[Irwin(1959)]{Irwin1959}
Irwin, J. B., 1959, \aj, 64, 149

\bibitem[Kawaler (2010)]{Kawaler2010}
Kawaler, S. D., 2010, AN, 331, 1020

\bibitem[Kilkenny et al.(1997)]{Kilkenny et al.1997}
Kilkenny, D., O'Donoghue, D., Koen, C., Stobie, R. S., \& Chen, A., 1997, \mnras, 287, 867

\bibitem[Kilkenny et al.(2016)]{Kilkenny et al.2016}
Kilkenny, D., Worters, H. L., O'Donoghue, D., Koen, C., Koen, T. Hambly, N. MacGillivray, H. \& Stobie, R. S., 2016, \mnras, 459, 4343

\bibitem[Koen (1998)]{Koen1998}
Koen, C., 1998, MNRAS, 300, 567

\bibitem[{{Kostov} {et~al.}(2016){Kostov}, {Moore}, {Tamayo}, {Jayawardhana},
  \& {Rinehart}}]{2016ApJ...832..183K}
{Kostov}, V.~B., {Moore}, K., {Tamayo}, D., {Jayawardhana}, R., \& {Rinehart},
  S.~A. 2016, \apj, 832, 183

\bibitem[{{Kupfer} {et~al.}(2015){Kupfer}, {Geier}, {Heber}, {{\O}stensen},
  {Barlow}, {Maxted}, {Heuser}, {Schaffenroth}, \&
  {G{\"a}nsicke}}]{2015A&A...576A..44K}
{Kupfer}, T., {Geier}, S., {Heber}, U., {et~al.} 2015, \aap, 576, A44

\bibitem[Lenz(2004)]{Lenz2004}
Lenz, P., 2004, Communications in Asteroseismology, 144

\bibitem[Lynas-Gray(2013)]{Lynas-Gray2013}
Lynas-Gray, A. E. 2013, in Astronomical Society of the Pacific Conference Series, Vol. 479, Progress in Physics of the Sun and Stars: A New Era in Helio- and Asteroseismology, ed. H. Shibahashi, \& A. E. Lynas-Gray, 273

\bibitem[Lutz(2011)]{Lutz2011}
Lutz, R., 2011, Ph.D thesis, Georg-August-Universit\"{a}t 

\bibitem[Mullally et al.(2008)]{Mullally et al.2008}
Mullally, Fergal, Winget, D. E., Degennaro, S., Jeffery, E., Thompson, S. E., Chandler, D., \& Kepler, S. O., 2008, \apj, 676, 573

\bibitem[{{N{\'e}meth} {et~al.}(2012){N{\'e}meth}, {Kawka}, \&
{Vennes}}]{2012MNRAS.427.2180N}
{N{\'e}meth}, P., {Kawka}, A., \& {Vennes}, S. 2012, \mnras, 427, 2180

\bibitem[O'Donoghue, Koen \& Kilkenny(1996)]{O'Donoghue et al.1996}
O'Donoghue, D., Koen C., \& Kilkenny D., 1996, \mnras, 278, 1075

\bibitem[O'Donoghue et al.(1997)]{O'Donoghue et al.1997}
O'Donoghue, D., Lynas-Gray, A. E., Kilkenny, D., Stobie, R. S., \& Koen, C., 1997, \mnras, 285, 657

\bibitem[Oreiro et al.(2004)]{Oreiro et al.2004}
Oreiro, R., Ulla, A., P\'{e}rez Hern\'{a}ndez, F., {\O}stensen, R., Rodr\'{i}guez L\'{o}pez, C., \& MacDonald, J., 2004, \aap, 418, 243.

\bibitem[{\O}stensen et al.(2001)]{Ostensen et al.2001}
{\O}stensen, R., Solheim, J.-E., Heber, U., Silvotti, R., Dreizler, S., \& Edelmann, H., 2001, \aap, 368, 175

\bibitem[{\O}stensen et al.(2010)]{Ostensen et al.2010}
{\O}stensen, R., Oreiro, R., Solheim, J. -E., Heber, U., et al., 2010, \aap, 513, 6

\bibitem[{{{\O}stensen} \& {Van Winckel}(2012)}]{2012ASPC..452..163O}
{{\O}stensen}, R.~H., \& {Van Winckel}, H. 2012, in Astronomical Society of the
  Pacific Conference Series, Vol. 452, Fifth Meeting on Hot Subdwarf Stars and
  Related Objects, ed. D.~{Kilkenny}, C.~S. {Jeffery}, \& C.~{Koen}, 163

\bibitem[Otani(2015)]{Otani2015}
Otani, T., 2015, Ph.D thesis, Florida Institute of Technology

\bibitem[Paparo, Szeidl \& Mandy(1988)]{Paparo et al.1988}
Paparo, M., Szeidl, B., \& Mahdy, H. A., 1988, \apss, 149, 73

\bibitem[{{Paxton} {et~al.}(2011){Paxton}, {Bildsten}, {Dotter}, {Herwig},
  {Lesaffre}, \& {Timmes}}]{2011ApJS..192....3P}
{Paxton}, B., {Bildsten}, L., {Dotter}, A., {et~al.} 2011, \apjs, 192, 3

\bibitem[{{Paxton} {et~al.}(2013){Paxton}, {Cantiello}, {Arras}, {Bildsten},
  {Brown}, {Dotter}, {Mankovich}, {Montgomery}, {Stello}, {Timmes}, \&
  {Townsend}}]{2013ApJS..208....4P}
{Paxton}, B., {Cantiello}, M., {Arras}, P., {et~al.} 2013, \apjs, 208, 4

\bibitem[{{Paxton} {et~al.}(2015){Paxton}, {Marchant}, {Schwab}, {Bauer},
  {Bildsten}, {Cantiello}, {Dessart}, {Farmer}, {Hu}, {Langer}, {Townsend},
  {Townsley}, \& {Timmes}}]{2015ApJS..220...15P}
{Paxton}, B., {Marchant}, P., {Schwab}, J., {et~al.} 2015, \apjs, 220, 15

\bibitem[Press et al.(1992)]{Press et al.1992}
Press W. H., Teukolsky S. A., Vetterling W. T., Flannery B. P., 1992, Numerical Recipes (Cambridge: Cambridge Univ. Press), Chapter 15

\bibitem[Qian et al.(2009)]{Qian et al.2009}
Qian, S. B. et al., 2009, \apj, 695, 163

\bibitem[Randall et al.(2006)]{Randall et al.2006}
Randall, S. K., Fontaine, G., Charpinet, S., Lynas-Gray, A. E., Lopes, I. P., O'Toole, S. J., \& Brassard, P., 2006, \apj, 648, 637

\bibitem[Randall et al. (2009)]{Randall et al.2009}
Randall, S. K., Van Grootel, V., Fontaine, G., Charpinet, S., \& Brassard, P. 2009, \aap, 507, 911

\bibitem[Saffer et al.(1994)]{Saffer et al.1994}
Saffer, R. A., Bergeron, P., Koester, D., \& Liebert, J., 1994, \apj, 533, 984

\bibitem[Schechter, Mateo, \& Saha(1993)]{Schechter et al.1993}
Schechter, P. L., Mateo, M., \& Saha, A., 1993, \pasp, 105, 1342

\bibitem[Schmidt-Kaler(1982)]{Schmidt-Kaler1982}
Schmidt-Kaler, Th., 1982, $Landolt-Bornstein New Series$, Vol. 2b, Springer Verlan, New York

\bibitem[Schuh et al.(2006)]{Schuh et al.2006}
Schuh, S., Huber, J., Dreizler, S., Heber, U., O'Toole, S. J., Green, E. M., \& Fontaine, G., 2006, \aap, 445, 31

\bibitem[Silvotti et al.(2007)]{Silvotti et al.2007}
Silvotti, R. et al., 2007, \nat, 449, 189

\bibitem[Silvotti et al.(2011)]{Silvotti et al.2011}
Silvotti, R., Szab\'{o}, R., Degroote, P., {\O}stensen, R. H., \& Schuh, S., 2011, AIP Conf. Proc., 1331, 133

\bibitem[Silvotti et al.(2014)]{Silvotti et al.2014}
Silvotti, R., {\O}stensen, R., Telting, J., \& Lovis, C., 2014, in Van Grootel, V., Green, E., Fontaine, G., \& Charpinet, S., eds, ASP Conf. Ser., Vol. 481, 6th Meeting on Hot Subdwarf Stars and Related Objects., Astron. Soc. Pac., San Francisco, p. 13

\bibitem[Stobie et al.(1997)]{Stobie et al.1997}
Stobie, R. S. et al., 1997, MNRAS, 287, 848

\bibitem[Sullivan et al.(2008)]{Sullivan et al.2008}
Sullivan, D. J., Metcalfe, T. S., O'Donoghue, D., Winget, D. E., et al., 2008, \mnras, 387, 137.

\bibitem[Tody(1986)]{Tody1986}
Tody, D., 1986, in D. L. Crawford, ed. Proc. SPIE Instrumentation in Astronomy VI, 627, 733

\bibitem[Tody(1993)]{Tody1993}
Tody, D., 1993, in eds. R.J. Hanisch, R. J. V. Brissenden, and J. Barnes, eds.  ASP Conf. Ser., Vol. 52, Astronomical Data Analysis Software and System II, 173

\bibitem[{{Veras} {et~al.}(2017){Veras}, {Georgakarakos}, {Dobbs-Dixon}, \&
  {G{\"a}nsicke}}]{2017MNRAS.465.2053V}
{Veras}, D., {Georgakarakos}, N., {Dobbs-Dixon}, I., \& {G{\"a}nsicke}, B.~T.
  2017, \mnras, 465, 2053

\bibitem[{{Vos} {et~al.}(2015){Vos}, {{\O}stensen}, {Marchant}, \& {Van
  Winckel}}]{2015A&A...579A..49V}
{Vos}, J., {{\O}stensen}, R.~H., {Marchant}, P., \& {Van Winckel}, H. 2015,
  \aap, 579, A49
  
  \bibitem[Vos et al.(2017)]{Vos2017}
Vos, J., {N{\'e}meth}, P. {Vu{\u{c}}kovi{\'c}}, M.,  {{\O}stensen}, R. \& Parsons, S., 2017, MNRAS, 473, 693
  
 \bibitem[{{V{\"o}lschow} {et~al.}(2016){V{\"o}lschow}, {Schleicher},
  {Perdelwitz}, \& {Banerjee}}]{2016A&A...587A..34V}
{V{\"o}lschow}, M., {Schleicher}, D.~R.~G., {Perdelwitz}, V., \& {Banerjee}, R.
  2016, \aap, 587, A34
  
\bibitem[Wade et al.(2014)]{Wade et al.2014}
Wade R., Barlow, B., Liss. S., \& Stark, M. 2014, in Van Grootel, V., Green, E., Fontaine, G., \& Charpinet, S., eds, ASP Conf. Ser., Vol. 481, 6th Meeting on Hot Subdwarf Stars and Related Objects. Astron. Soc. Pac., San Francisco, p. 311

\bibitem[Winget \& Kepler(2008)]{Winget2008}
Winget, D. E., \& Kepler, S. O., 2008, \araa, 46, 157

\bibitem[{{Xiong} {et~al.}(2017){Xiong}, {Chen}, {Podsiadlowski}, {Li}, \&
  {Han}}]{2017A&A...599A..54X}
  {Xiong}, H., {Chen}, X., {Podsiadlowski}, P., {Li}, Y., \& {Han}, Z. 2017,
  \aap, 599, A54
  
\bibitem[{{Zorotovic} \& {Schreiber}(2013)}]{2013A&A...549A..95Z}
{Zorotovic}, M., \& {Schreiber}, M.~R. 2013, \aap, 549, A95

\end{thebibliography}
\end{document}